\documentclass[a4paper,12pt]{article}
\usepackage[pdftex]{graphicx}
\usepackage{amssymb,latexsym}
\usepackage{latexsym,amsfonts,amsmath,amscd,epsfig}
\usepackage{amssymb,amsxtra,mathtools}
\usepackage{url}
\usepackage{pdflscape}
\usepackage[english]{babel}
\newcommand{\bec}{\begin{center}}
\newcommand{\ec}{\end{center}}
\newcommand{\bee}{\begin{equation}}
\newcommand{\ee}{\end{equation}}

\title{Computer modeling of properties of Kaluza-Klein particles and their searches at the LHC}

\author{T.V. Obikhod\thanks{E-mail: obikhod@kinr.kiev.ua}, I.A. Petrenko\\
\small\emph{Institute for Nuclear Research, National Academy of Science of Ukraine} \\ 
\small\emph{47, prosp. Nauki, Kiev, 03028, Ukraine}}

\date{\small\today}
\begin{document}
\maketitle

\abstract{The Standard Model problems lead to the new theories of extra dimensions: Randall-Sundrum model, Arkani-Hamed-Dimopoulos-Dvali model and 
TeV$^{-1}$ model. In the framework of these models with the help of computer program Pythia8.2 were calculated the production cross sections for Kaluza-Klein particles at various energies at the LHC. The generation of monojet events from scalar graviton emission was considered for number of extra dimensions, n=2, 4, 6, for the energy at the LHC 14 TeV. Also are studied the graviton production processes through the gluon-gluon, quark-gluon and quark-quark fusion processes and found some periodicity in the behavior of the graviton mass spectrum.  Within Randall-Sundrum scenario were calculated $\sigma\times$ Br for production process of massive graviton, gg $\rightarrow$ $G^{*}$, and the most probable processes of graviton decay at 13 TeV, 14 TeV and 100 TeV.} 

\newpage
\section{Introduction}
The problems with theoretical explanation of vacuum energy as well as dark energy, dark matter and cosmological constant problems are only the tip of the iceberg of problems in the modern theoretical physics. Some of them are:\\
$\bullet$ ordinary matter accounting for about 5$\%$ of mass energy in the Universe and no dark matter candidate in the Standard Model (SM); \\
$\bullet$ hierarchy problem;\\
$\bullet$ fine tuning of SM Higgs mass; \\
$\bullet$ no explanation for fermion masses and mixings and three family structure; \\
$\bullet$ 	unification of strong, electro-weak, and gravitational forces; \\
$\bullet$ compositeness of leptons and quarks \\
and so on. It is an experimental fact that there is something we can't explain within the SM.

	As is known, vacuum is produced in the processes of phase transitions in Early Universe and the space-time structure of the physical vacuum exhibits the connection to the structure formation in our Universe. So, the understanding of Universe formation is deeply connected with the conception of the space-time. According to hierarchy formula \cite{1.}, Plank energy can be reduced to the energy of about 10 TeV that is achieved at the LHC. So, the phenomena of the Universe formation at the early stages and the accompanying processes of particle physics could be studied at modern colliders. In spite of the fact that no new physics beyond the SM is discovered at the LHC at 13 TeV, the planned upgrading of the LHC to high luminosities and energies up to 100 TeV gives the possibility for the discovery of new physics. papers were about . Among such searches for new physics the most popular are the experimental searches for the Kaluza Klein (KK) particles. 
	
	Historically, KK theory appeared as the unification of gravitational and electromagnetic interactions due to the proposition of a fifth dimension in addition to the usual four-dimensional  space-time \cite{2.}, that leads to the consideration of the concept of deformation of Riemannian geometry defined by extrinsic curvature of the space-time. The conclusions of this result are based, in particular, on the five-dimensional space from the paper \cite{3.}. N. Arkani-Hamed et al. proposed the solution to the hierarchy problem on the basis of the existence of new compact spatial dimensions. KK excitations in this extra dimensional space through the combined effect of all the gravitons became observable. 
	
	Today, the idea of additional space as the instrumentation of the unification of all four interactions is of interest not only in theoretical physics 
	\cite{4., 5.}, but in experimental searches at the LHC for  exotic matter  that deviates from normal matter  \cite{6.}. 
	
	Our paper is devoted to the searches for KK particles in three models of extra dimensions: Arkani-Hamed-Dimopoulos-Dvali (ADD) model, \cite{4.}, Randall-Sundrum (RS) model \cite{5.} and TeV$^{-1}$ model \cite{7.}. Using computer program Pythia8.2 \cite{8.} within these three extra dimensional models, we have calculated:\\
$\bullet$  the production cross section of KK modes of massive gravitons and gauge bosons at energies from 14 TeV to planned 100 TeV; \\
$\bullet$  the graviton mass spectrum for three G jet emission processes: a) gg$\rightarrow Gg$, b) qg$\rightarrow Gq$, c) $q\overline{q}\rightarrow Gg$ at 14 TeV at the LHC;\\
$\bullet$ the graviton mass spectrum at 14 TeV for numbers of extra dimensions, n = 2, 4, 6;\\ 
$\bullet$  the production cross section of graviton, $gg \rightarrow G^{*}$, multiplied by branching ratios, Br$(G^{*} \rightarrow gg)$, Br$(G^{*} \rightarrow ll)$, Br$(G^{*} \rightarrow hh)$ of the most probable processes of decay within RS model at 13 TeV, 14 TeV, 100 TeV.

\section{Models of extra dimensions}
In this section we'll observe three models of extra dimensions, ADD, RS and 
TeV$^{-1}$, which are the base for our further calculations of KK particle properties. 
In the framework of M-theory \cite{9.}, the metric of ADD model: 
\[G_{MN}(x, y)=\eta_{MN}+\frac{2}{M^{1+n/2}}h_{MN}(x, y),\]
where $G_{MN}(x, y)$ is the metric of (4 + n)-dimensional space-time with compact extra dimensions where the gravitational field propagates and the SM localized on a 3-brane embedded into the (4 + n)-dimensional space-time, $\eta_{MN}$ is (4 + n)-dimensional Minkowski background and $h_{MN}(x,y)$ is the deviation of Minkowski metrics, $M$ is the fundamental mass scale and n - the number of extra dimensions. Masses of KK modes for ADD model are given by 
\[m_n=\frac{1}{R}\sqrt{n_1^2+n_2^2+\ldots + n_d^2}=\frac{|n|}{R}\]
Five-dimensional metric with one extra dimension compactified to the orbifold, S$^1$/Z$_2$, obtained by L. Randall and R. Sundrum \cite{5.}, is with non-factorizable geometry 
\[ds^2=e^{-2\sigma(y)}\eta_{\mu\nu}dx^{\mu}dx^{\nu}+dy^2\ .\]
Two 3-branes are located at points y=0 and y= $\pi R$ of the orbifold with R - radius of S$^1$. $x^{\mu}, \mu = 0, 1, 2, 3$ and $\eta_{\mu\nu}$  are four-dimensional coordinates and Minkowski metrics, the function $\sigma (y)$ inside the interval $-\pi R < y < \pi R$  is equal to $\sigma (y)=k|y|$. ($k>0$ - dimensional parameter). In Fig. 1 is presented a non factorizable geometry with one spatial extra dimension as a line segment between two four dimensional branes, known as Planck and TeV brane.  
\bec
{\includegraphics[width=0.35\textwidth]{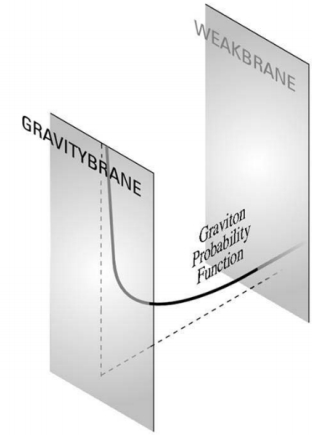}}\\
\emph{{Fig.1.}} {\emph{RS theory presented by the Gravity and Weak branes 
as the 4 dimensional boundaries of the extra dimension, from [10].}}\\
\ec
Masses of KK particles for RS model are given by 
\[m_n=\beta_nke^{-\pi k R},\]
$\beta_n=3.83, 7.02, 10.17, 13.32, ...$ for $n=1, 2,  3, 4, ... . $ 
The metric of TeV$^{-1}$ model for ten-dimensional string theory is determined by the conditions on Calabi-Yau manifold: Ricci-flatness of metric, vanishing first Chern class and SU(n) holonomy. Low-energy effective action has scale much smaller than Planck mass that related to an internal compactification radius. In these scenarios, the SM fields as well as Z$_{KK}$ and W$_{KK}$ resonances are allowed to propagate in the bulk, but gravity is not included in the model. Masses of KK modes for TeV$^{-1}$ model are given by 
\[m_n=(m_0^2+n\cdot n/R_c^2)^{1/2},\]
where $n=(n_1, n_2, ...)$ labels KK excitation levels, $m_0$ is the mass of gauge zero-mode, which corresponds to SM gauge field.

	The advantage of the presented models of extra dimensions lies in the possibility of the observation of the physics of Planck scales, $M_{Pl}$,  at the energies achievable at modern colliders, $M$, due to the presence of extra dimensions, n. This result is possible due to the hierarchy formulas for the presented models:\\
ADD model: $M_{Pl}^2\sim R^n M^{2+n}$;\\
RS model: $M_{Pl}^2=\frac{M^3}{k}(e^{2k\pi R}-1)$;\\
TeV$^{-1}$ model: hierarchy formula is determined by the low-energy effective action.

\section{Results of computer modeling for properties of KK particles}
RS resonances, connected with production of KK graviton, G$^*$, are described in \cite{11.}. The production of narrow graviton resonances in the TeV range at the LHC as well as the decays into fermion and boson pairs was studied in this paper. For the discovering of graviton resonance, G$^*$, was used the parton showering formalism, which agrees with the NLO matrix element calculations. The partonic subprocesses are demonstrated in Fig.2.

\bec
{\includegraphics[width=0.55\textwidth]{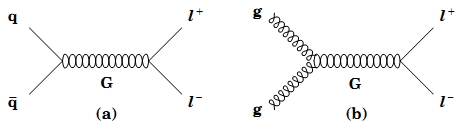}}\\
\emph{{Fig.2.}} {\emph{Processes for the graviton resonance production through a) quark-quark b) gluon-gluon fusion, from \cite{11.}.}}\\
\ec

The ADD graviton emission and virtual graviton exchange processes are described in \cite{12.}. Within model with large extra dimensions (LED) were considered the processes that could give rise to new phenomena at LHC due to emission or exchange of gravitons. The implementation of these processes in the Monte Carlo generator Pythia8.2 was presented in this paper. The considered processes are connected with mono-jet, di-photon and di-lepton final states. It is also possible to generate monojet events from scalar graviton emission as described in \cite{13.}.

	TeV$^{-1}$ Sized Extra Dimensional KK production processes involve the production of electroweak KK gauge bosons, Z$_{KK}$ and W$_{KK}$, in one TeV$^{-1}$ sized extra dimension. The processes are described in \cite{14.} and allow the specification of final states. In this article the observation of a certain KK hard process in pp interactions at the LHC was presented within the S$^1$/Z$_2$, TeV$^{-1}$ extra dimensional theoretical framework. The analytic form for the hard process cross section has been calculated and has been implemented within the Pythia8.2 Monte Carlo generator. 

\subsection{The production cross section of KK particles at different energies}
With the help of computer program Pythia8.2 for three models of extra dimensions we have calculated the production cross sections of KK particles at energies varied from 14 to 100 TeV. The results are presented in Fig.3.

\bec
{\includegraphics[width=0.82\textwidth]{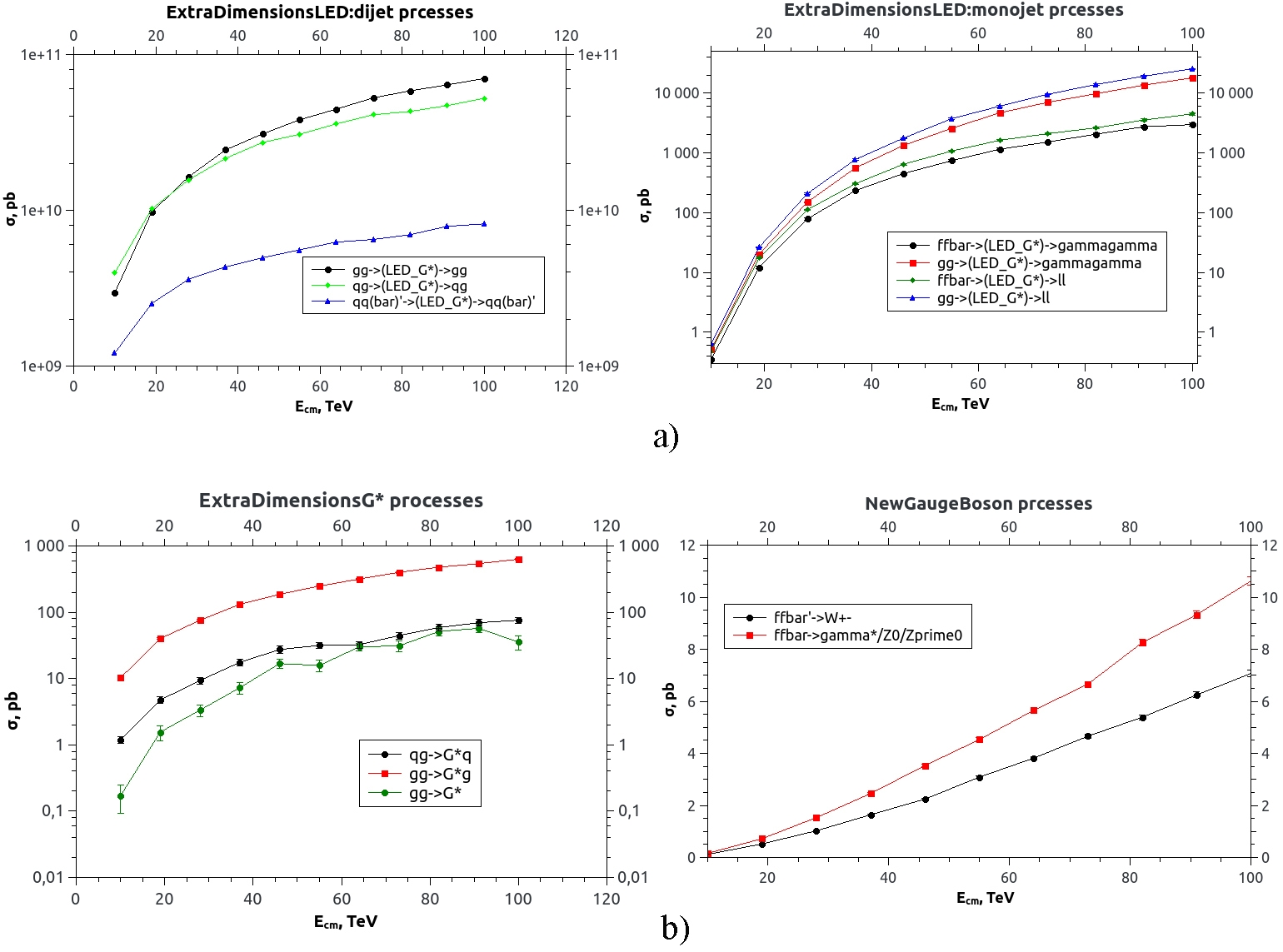}}\\
\emph{{Fig.3.}} {\emph{The production cross section at the center of mass energies varied from 14 to 100 TeV for 
a) LED model: left – dijet final state, right – monojet final state; 
b) left – RS model, right – TeV$^{-1}$ model.}}\\
\ec

From Fig. 3 a) we can see the significant predominance of production for dijet final states above monojet final states within LED model. Fig. 3 b) shows that RS KK particles are produced at much higher values than Z$_{KK}$ and W$_{KK}$ bosons in TeV$^{-1}$ extra dimensional model. 

	Within TeV$^{-1}$ model with parameters n=10 ,m$^*$=2-10 TeV we calculated the production cross sections at 100 TeV, 60 TeV and 20 TeV at the center of mass energies as a function of KK mass, M$_{Z_{KK}}$, presented in Fig. 4. As the decay of Z$_{KK}$ to muon pair is the dominant one, we decided to calculate namely this process of KK particle decay.  

\bec
{\includegraphics[width=0.65\textwidth]{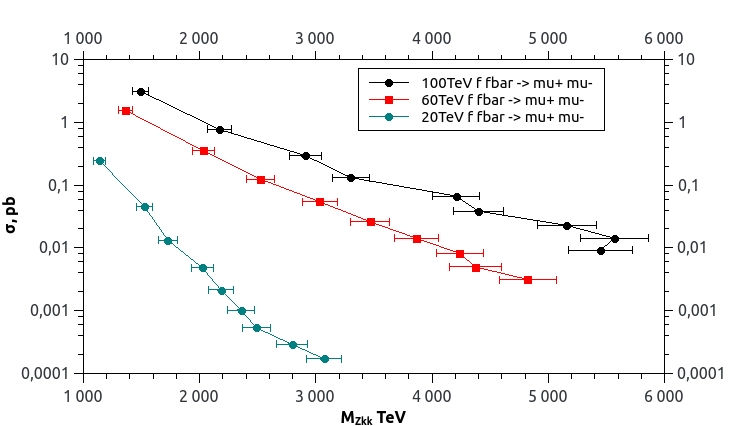}}\\
\emph{{Fig.4.}} {\emph{Production cross sections at 100 TeV, 60 TeV and 20 TeV at the center of mass energies as a function of KK mass for TeV$^{-1}$ model.}}\\
\ec

From Fig. 4 we can see the sharp drop in the curve for 20 TeV compared with other curves at 60 TeV and 100 TeV at the center of mass energies. This result emphasizes the most important result for the further searches of new physics at high energies. As the last two curves (60 TeV and 100 TeV) are almost parallel, it is preferable to search for new phenomena at energies up to 30 TeV, when the production cross sections of KK particles can be varied in the wide range. 

\subsection{The graviton mass spectrum at 14 TeV}

We'll apply large-extra-dimensional (ADD-type) models in production processes for virtual extra-dimensional scalars of graviscalar type. With the help of Pythia8.2 it is possible to generate monojet events from scalar graviton emission as described in \cite{13.}. 

	The group of lowest-order G jet emission processes, within monojet model was considered with the following parameters:\\ 
ExtraDimensionsLED:n = 6\\
ExtraDimensionsLED:M = 10000.\\
Three graviton mass spectrum for G jet emission processes, gg$\rightarrow Gg$, qg$\rightarrow Gq$, $q\overline{q}\rightarrow Gg$ at the center of mass energies of 14 TeV are presented in Fig. 5.

\bec
{\includegraphics[width=0.45\textwidth]{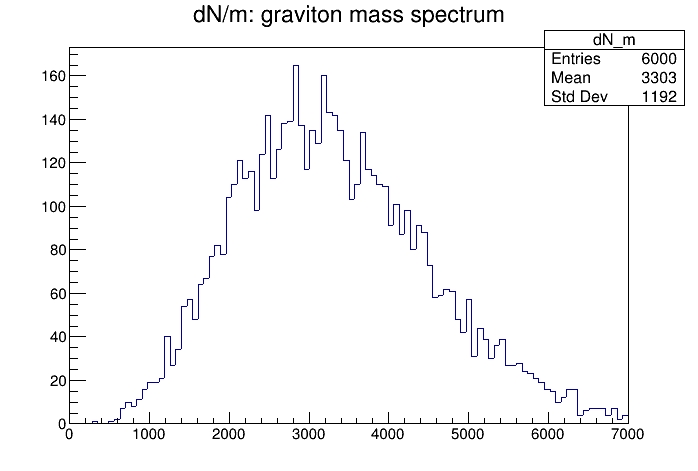}
\bec
a)
\ec}
{\includegraphics[width=0.45\textwidth]{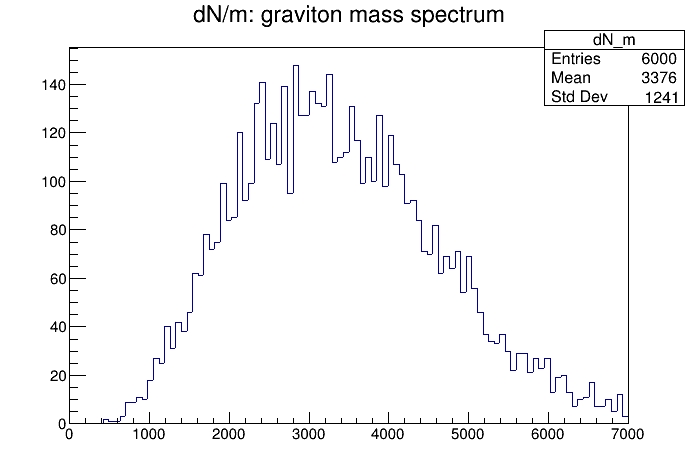}
\bec
b)
\ec}
{\includegraphics[width=0.45\textwidth]{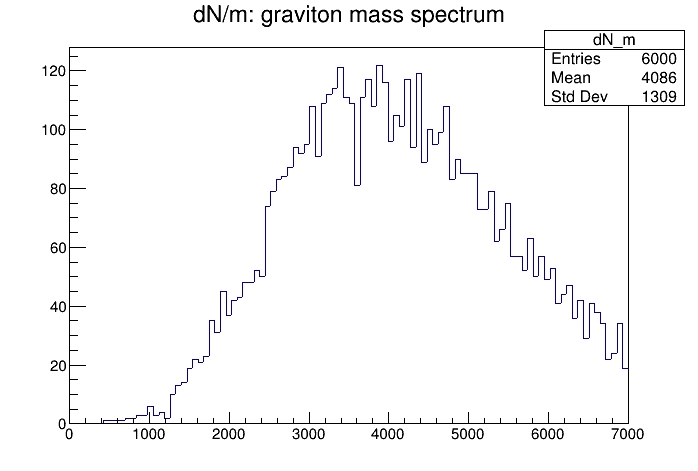}
\bec
c)
\ec}
\emph{{Fig.5.}} {\emph{Graviton mass spectrum within monojet LED model for three G jet emission processes a) gg$\rightarrow Gg$, b) qg$\rightarrow Gq$, c) $q\overline{q}\rightarrow Gg$ at the center of mass energies of 14 TeV.}}\\
\ec
From Fig. 5, the peak of the graviton mass spectrum distribution is viewed depending on the process of monojet emission. Although this dependence is insignificant, nevertheless, for the process $q\overline{q}\rightarrow Gg$ of monojet emission it is shifted by almost 1 TeV.

	As LED model depends on the number of extra dimensions, n, it was important to study the graviton mass spectrum distributions for n=2, 4, 6. The results of our calculations of G jet emission process, gg$\rightarrow Gg$ at the center of mass energies of 14 TeV is presented in Fig. 6

\bec
{\includegraphics[width=0.45\textwidth]{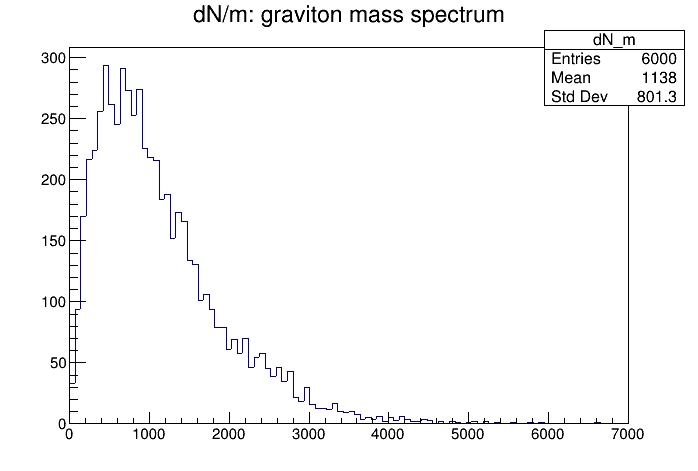}
\bec
a)
\ec}
{\includegraphics[width=0.45\textwidth]{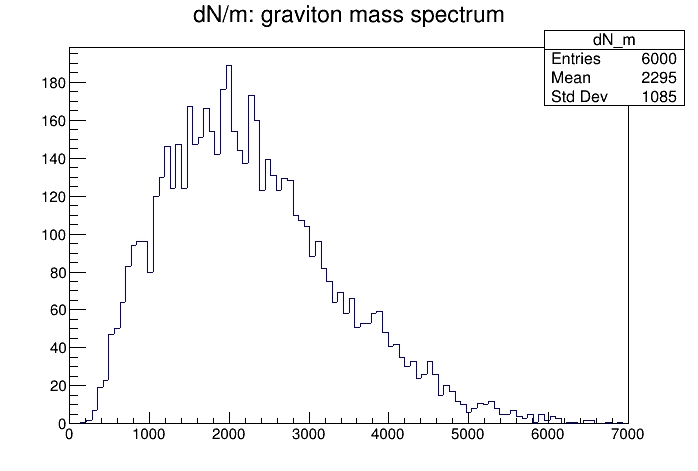}
\bec
b)
\ec}
{\includegraphics[width=0.45\textwidth]{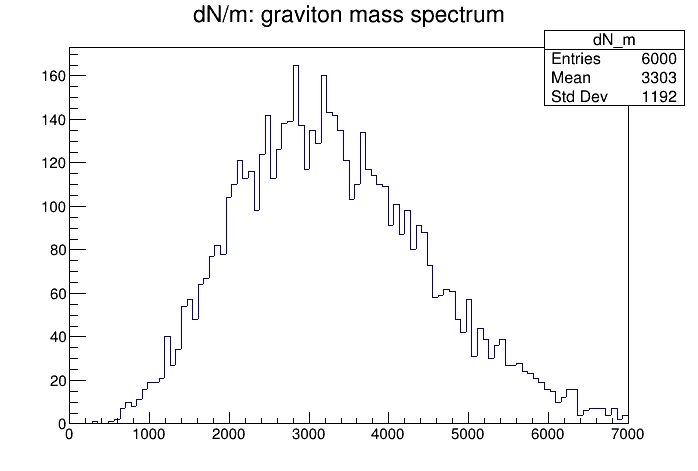}
\bec
c)
\ec}
\emph{{Fig.6.}} {\emph{Graviton mass spectrum within monojet LED model for G jet emission process gg$\rightarrow Gg$ for a) n=2, b) n=4, c) n=6 at the center of mass energies of 14 TeV.}}\\
\ec
From Fig. 6 is seen substantial dependence of G jet emission process in the LED model on the number of extra dimensions. Moreover, we can see a clear periodicity of dependence, when peaks are shifted by 1 TeV with an increase of the number of extra dimensions by 2. 

\subsection{The production cross section of graviton emission multiplied by branching ratio in RS model}
As is known, the discovery of a Higgs boson, $h$, at the LHC motivates the searches for physics beyond SM in channels involving Higgs boson. Higgs pair production is predicted by RS model with KK graviton, G$^*_{KK}$, emission that may decay to a pair of Higgs bosons. Such experimental searches were performed by ATLAS Collaboration, \cite{15.} at 13 TeV and presented in Fig. 7. 

\bec
{\includegraphics[width=0.78\textwidth]{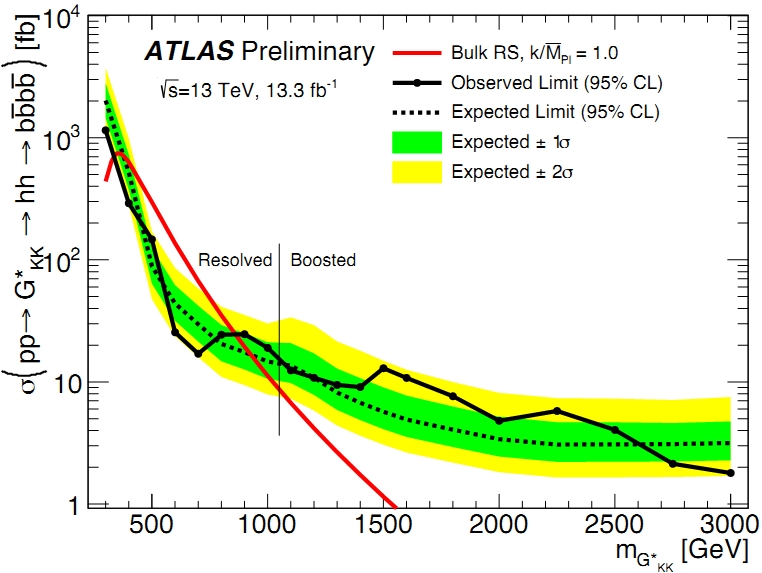}}\\
\emph{{Fig.7.}} {\emph{The expected and observed upper limit for 
pp$\rightarrow  G^*_{KK}\rightarrow hh\rightarrow b\overline{b} b\overline{b}$
in the bulk RS model with k/M = 1 at the 95$\%$ confidence level.}}
\ec
\newpage

	From \cite{16.} are taken branching fractions of graviton decay to SM particles, Fig. 8. The predictions for decay were updated using state-of-the-art computation tools with the highest branching fraction to di-jet final states.
\bec
{\includegraphics[width=0.55\textwidth]{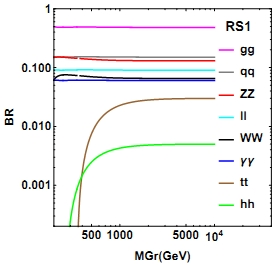}}\\
\emph{{Fig.8.}} {\emph{Branching fractions of graviton, G$^*$, from \cite{16.}.}}\\
\ec

	We will consider three processes of graviton decay, 
G$^*\rightarrow gg$; G$^*\rightarrow ll$; G$^*\rightarrow hh$, for further 
$\sigma\times Br$ calculations within RS scenario for graviton production process gg$\rightarrow G^*$ at 13 TeV, 14 TeV and 100 TeV at the center of mass energies. In Fig. 9 are presented our calculations performed with the help of Pythia8.2, with parameters k/M = 1 and 0.1. 
\bec
{\includegraphics[width=0.59\textwidth]{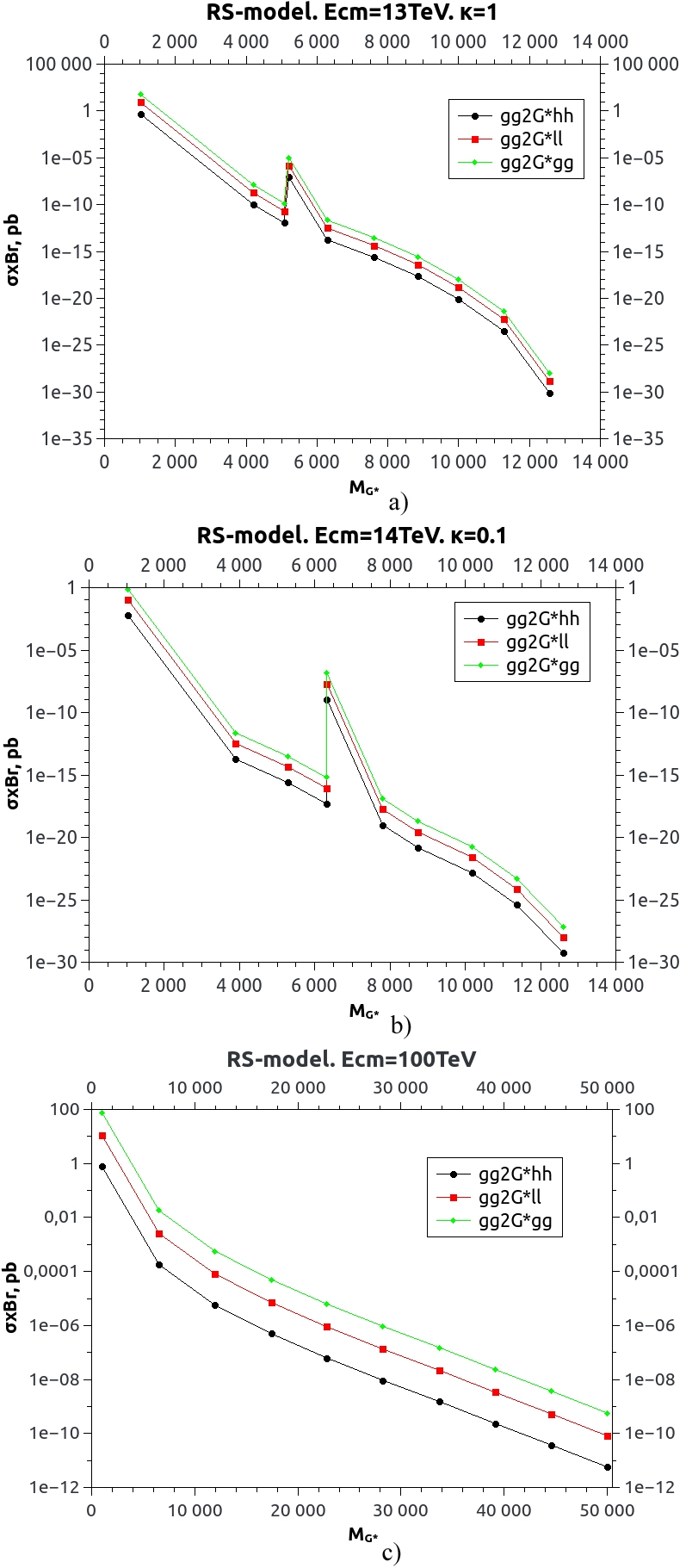}}\\
\emph{{Fig.9.}} {\emph{$\sigma\times Br$ for graviton production and decay as the function of graviton mass, $M_{G^*}$, at a)13 TeV, k/M = 1; b) 14 TeV, k/M = 0.1; c) 100 TeV, k/M = 0.1.}}
\ec
From Fig. 9 we see the dependence of the $\sigma\times Br$ calculations on the parameter k/M and on the energy at the center of mass. Comparison of Fig. 9 a) and b) shows that the resonance peak shifts from 5 TeV to 7 TeV with the increasing of energy at the collider. In the case of Fig. 9 c), we can see that there is no peak at 100 TeV and that the cross section for the formation of the resonance as a function of energy is observed to decrease.

\section{Conclusion}
The modern high energy physics is connected with experimental searches of new physics beyond the SM. These searches are connected not only with new possibilities of modern accelerating technics but also with problems of SM physics. The SM problems are not only of theoretical character but also rather of experimental one, which is confirmed by modern experiments on the Higgs boson. Our work is dedicated to the studying of the properties of the new particles predicted by the theories of extra dimensions. Within three models, ADD, RS and TeV$^{-1}$ we have calculated the production cross sections of massive graviton formation as well as the production of KK modes of gauge bosons depending on the energies of the modern and future colliders. Within LED model the behavior of graviton mass spectrum for G jet emission processes for different numbers of extra dimension, n=2, 4, 6 was studied and clear periodicity of peaks shifted by 1 TeV was seen with an increase in the number of extra dimensions by 2. Also was investigated the graviton mass spectrum for three G jet emission processes: a) gg$\rightarrow Gg$, b) qg$\rightarrow Gq$, c) $q\overline{q}\rightarrow Gg$ at 14 TeV at the LHC. The experimental searches for KK graviton emission and decay to a pair of Higgs bosons, performed by ATLAS Collaboration at 13 TeV stimulated us to perform calculations at different parameters and energies within RS model. Our calculations of $\sigma\times Br$ shows that the resonance peak shifts from 5 TeV to 7 TeV with the increasing of energy at the colliders from 13 TeV to 14 TeV as well as the absence of peak at energy of 100 TeV at the center of mass energies.

\label{page-last} 
\label{last-page}
\end{document}